\documentclass[aps,prd,groupedaddress]{revtex4-2}
\usepackage{amsmath,amssymb,amsfonts}
\usepackage{graphicx}
\usepackage[english]{babel}
\usepackage{physics}
\usepackage[dvipsnames]{xcolor}
\usepackage{aas_macros}
\usepackage{soul}

\usepackage{hyperref}
\hypersetup{
    colorlinks=true,
    linkcolor=blue,
    citecolor=blue,
    filecolor=magenta,      
    urlcolor=blue,
    pdftitle={Schroer+,2025},
    pdfpagemode=FullScreen,
    }

\usepackage{orcidlink}

\usepackage{lineno}
\begin{document}
\title{A Critical Examination of the Nested Leaky Box Model for Galactic Cosmic Ray Transport}

\author{Benedikt Schroer\,
\orcidlink{0000-0002-4273-9896}}
\email[]{bschroer@uchicago.edu}
\affiliation{Department of Astronomy \& Astrophysics, University of Chicago, 5640 S Ellis Ave, Chicago, IL 60637, USA}

\author{Carmelo Evoli\,\orcidlink{0000-0002-6023-5253}}
\email[]{carmelo.evoli@gssi.it}
\affiliation{Gran Sasso Science Institute (GSSI), Viale Francesco Crispi 7, 67100 L’Aquila, Italy}
\affiliation{INFN-Laboratori Nazionali del Gran Sasso (LNGS), via G. Acitelli 22, 67100 Assergi (AQ), Italy}

\author{Pasquale Blasi\,
\orcidlink{0000-0003-2480-599X}}
\email[]{pasquale.blasi@gssi.it}
\affiliation{Gran Sasso Science Institute (GSSI), Viale Francesco Crispi 7, 67100 L’Aquila, Italy}
\affiliation{INFN-Laboratori Nazionali del Gran Sasso (LNGS), via G. Acitelli 22, 67100 Assergi (AQ), Italy}

\newcommand{\CE}[1]{{\color{red}#1}}
\newcommand{\bs}[1]{{\color{orange} #1}}

\date{\today}

\begin{abstract}
We revisit the predictions of the nested leaky box model in detail, in terms of both primary cosmic-ray spectra, spectra of stable and unstable nuclei and antimatter production (positrons and antiprotons). We conclude that the model is in direct conflict with several observational facts and at least in its vanilla version should be considered as ruled out by current data. We also speculate on some possibly interesting implications of the idea that at least a fraction of Galactic grammage may not be accumulated in the journey of cosmic rays in the interstellar medium but rather inside the sources or in regions around them. These speculations will become increasingly more relevant with the higher precision data becoming available at high energies. 
\end{abstract}

\maketitle

\section{Introduction}
\label{sec:intro}

The discovery of the anomalous behavior of cosmic-ray (CR) positrons by PAMELA~\cite{PAMELA:2008gwm}, later extended by AMS-02~\cite{AMS:2013fma} to higher energies, led to an enormous effort in trying to clarify the nature of the anomaly. More than a decade later, it seems clear that pulsars are sizable contributors to the Galactic CR positron flux~\cite{Hooper:2008kg,Fermi-LAT:2009ppq,Delahaye:2010ji,Blasi:2010de,DiMauro:2014iia,Evoli:2020szd}, and such a contribution needs to be included in any prediction of the positron flux. On the other hand, it has been speculated that the rising positron fraction is a problem only in the context of the standard picture of CR transport, while one can construct alternative frameworks in which the observed positrons are purely secondary products of CR hadronic interactions~\cite{Blum:2013zsa,Lipari:2016vqk,Lipari:2021edz}. In fact, one such models, the nested leaky box model (NLBM), was first introduced as a possible explanation of the decreasing secondary-to-primary ratios of CR nuclear fluxes with energy, in terms of grammage accumulated near the sources rather than {\it en route} during Galactic propagation~\cite{Cowsik:1973yd,Cowsik:1975tj}. Interestingly, virtually the same model was recently applied to Galactic CRs and claimed to provide a natural explanation of B/C and similar secondary/primary ratios, as well as of positrons~\cite{Cowsik:2010zz} and antiprotons~\cite{Cowsik:2016wso,Cowsik:2022eik}. The idea behind the model is, in principle, rather simple: if CRs accumulate their grammage near the sources (in the so-called cocoons) in an energy dependent way, while transport on Galactic scales is energy independent, then at low energies per nucleon the B/C ratio is expected to reflect the energy dependence of the grammmage in the cocoons, while at large enough energies the B/C (and similar ratios of the fluxes of stable secondary and primary nuclei) should become energy independent. On the other hand, positrons and antiprotons are produced as secondary products of $pp$ interactions with inelasticity $\sim 0.05-0.1$.
As a result, most positrons and antiprotons observed at Earth are generated by collisions of high-energy primary nuclei during their Galactic transport, which is assumed to be energy independent. The NLBM posits that in order to fit the flatness of the ratio $e^+/\bar p$ with energy, positrons should not suffer severe losses at energies below $\sim 1$ TeV. The steepening in the electron plus positron flux at $\sim 1$ TeV~\cite{HESS:2008ibn,DAMPE:2017fbg,CALET:2023emo} can then be interpreted as the onset of losses (see also~\cite{Lipari:2022imm}). 
If confirmed, this picture would shake the pillars of the standard model of CR transport in the Galaxy, developed through decades of theoretical investigation. 

Recent developments in the field came from an apparently different direction: the discovery of TeV halos~\cite{HAWC:2017kbo,LHAASO:2021crt}, regions of high-energy gamma-ray emission around selected pulsar wind nebulae (PWNe), that has been interpreted as the result of a suppressed diffusivity in such regions~\cite{Lopez-Coto:2022igd}. If this phenomenon were sufficiently universal around Galactic CR sources, then it would be plausible that CRs may accumulate at least part of their grammage in trying to escape the regions of reduced diffusivity, the closest realization of the cocoons that are central to the NLBM. 

Given these findings we embarked on a detailed investigation of the phenomenological implications of the NLBM, based on the comparison between the predictions of the model and the data on the spectra of primary and secondary stable and unstable nuclei, as well as the spectra of CR positrons and antiprotons. The results of our investigation can be summarized as follows: 
1) since in the NLBM the hardening of the secondary/primary ratios~\cite{AMS:2021nhj,CALET:2022dta,DAMPE:2022vwu} is interpreted as the result of a transition from CR interactions during transport in the cocoons (energy dependent) to Galactic transport (independent of energy), no similar hardening should be showing in the spectrum of primary nuclei. This is in contradiction with the data of AMS-02 \cite{AMS:2015tnn,AMS:2015azc}, DAMPE~\cite{DAMPE:2019gys,DAMPE:2023pjt} and CALET~\cite{CALET:2019bmh,CALET:2023nif}, which show a similar phenomenon in the spectra of H and He nuclei (and in fact in virtually all other primary spectra). In order to accommodate this finding, the NLBM must introduce by hand these spectral breaks in the primary spectra, with no apparent physical justification. 
Furthermore, the hardening in the secondary nuclei is twice as large as that in the spectra of primaries and occurs at virtually the same rigidity, something that finds a natural explanation in the standard model~\cite{Evoli:2023kxd}. 
On the other hand, in the NLBM, this is purely by coincidence. 
2) As a consequence of the energy-independent transport in the Galaxy, the slope of the spectra of H and He nuclei at injection is the same as observed at Earth, which implies that the sources need to inject very steep spectra, somewhat at odds with the theoretical predictions of typical acceleration models~\cite{Caprioli:2023orv}. 
%3) Yet, the hardening in the secondary/primary ratios and in the spectra of primary nuclei occur at virtually the same rigidity. In the NLBM this is purely by coincidence. 
3) Our calculations show that the preliminary $^{10}$Be/$^{9}$Be ratio as measured by AMS-02~\cite{Dimiccoli:2023dbx} is incompatible with the predictions of the NLBM. The measurements of upcoming experiments, such as HELIX~\cite{HELIX:2023twg} will play a crucial role in assessing this point further. 
4) We find that the flat observed ratio $e^+/\bar p$, claimed to be one of the strongholds of the NLBM, is in fact not reproduced when the energy losses necessary to explain the steepening of the positron flux at $\sim 1$ TeV, are included in a self-consistent manner. 
5) Once the most recent cross sections for production of antiprotons are adopted, we show that the observed spectrum of antiprotons excludes an energy independent transport in the Galaxy, another pillar of the NLBM. 
6) Within the NLBM framework, the grammage required to reproduce B/C is incompatible with the grammage needed to explain antiprotons and positrons.
7) Finally, the rising positron fraction requires that in the NLBM the source spectrum of electrons is much stepper than that of protons, at odds with the rigidity dependent nature of CR acceleration processes~\cite{Cristofari:2021hbc}. 

The article is organized as follows: in Sect.~\ref{sec:model} we discuss the mathematical aspects and the assumptions of the NLBM and the reasons for its inception. In Sect.~\ref{sec:results} we illustrate the results of our calculations in terms of spectra of primary and secondary nuclei, unstable nuclei, positrons and antiprotons. Given the emphasis in the literature to the case of antiprotons having the same spectrum as primary protons, in Sect.~ \ref{sec:xsec} we discuss the implications of the cross section for antiproton production on the $\bar p/p$ ratio. Our conclusions are drawn in Sect.~\ref{sec:conclude}.

%%%%%%%%%% SECTION %%%%%%%%%%
\section{Review of the NLBM}
\label{sec:model}

In this section, we outline the main aspects and assumptions involved in the NLBM, for primary and secondary stable and unstable nuclei, for positrons and antiprotons.  

The NLBM envisions CRs diffusing in a hierarchy of two (or potentially more) nested zones. Each zone is characterized by distinct leakage rates,  with the dependence of leakage on cosmic-ray energy varying across the zones. Central to the NLBM is the assumption that the Galactic gas disk, modeled as a cylinder with a radius \( R_{\rm G} \approx 10 \) kpc and a half-height \( h \approx 100 \) pc, is uniformly populated with CR sources that accelerate nuclei following a nearly universal power-law spectrum. While this aspect is very similar to the standard model of CR transport, in the NLBM the same region is also where Galactic CR propagation occurs, namely there is no extended halo around the disc. 

Following acceleration, CRs undergo energy-dependent diffusion within localized regions surrounding sources, termed \emph{cocoons}. Within these cocoons, the escape rate of CRs, \( \tau_{\rm c}^{-1} \), increases with particle energy, leading to progressively shorter residence times for higher-energy particles. During their residence in the cocoons, CRs are primarily subject to spallation processes at lower energies, after which they escape into the broader interstellar medium. Leptons are assumed to not lose energy appreciably inside the cocoons. 

Upon leaving the cocoons, CRs fill the Galactic zone, continuing to diffuse and potentially engaging in additional interactions until they ultimately exit the Galaxy. 
In virtually all versions of the NLBM, it is posited that the Galactic residence time for CR nuclei, \( \tau_{\rm G} \), remains energy-independent up to several hundred TeV and small enough so that energy losses of leptons remain relatively small up to energies $\sim 1$ TeV~\cite{Cowsik:1973yd,Cowsik:1975tj,Cowsik:2010zz,Cowsik:2016wso}. This assumption is perceived as critical for explaining several observational phenomena, including the observed flattening of the boron-to-carbon (B/C) ratio at energies exceeding \( \sim \) TeV/n and the constant ratios of positrons-to-protons and antiprotons-to-protons at energies above approximately 60 GeV~\cite{AMS:2021nhj}. 

\subsection{Primary and Secondary Nuclei}

In the NLBM, the equilibrium density \( N_{{\rm c},i} \) of a CR species \( i \) (either primary or secondary in origin) within a cocoon is calculated based on the source emissivity \( Q_{{\rm c},i} \), the energy-dependent escape time \( \tau_{\rm c} \), the grammage \(\chi_{\rm c}=c n_{\rm c} \tau_{\rm c} m_p\), and the critical grammage \(\chi_i=\frac{m_p}{\sigma_i}\) for nucleus \( i \) due to inelastic collisions, as defined by the equation:
\begin{equation}\label{eq:cocoon}
N_{{\rm c},i}(E) = \left(1+\frac{\chi_{\rm c}}{\chi_i}\right)^{-1}\tau_{\rm c}(E) Q_{{\rm c},i}(E),
\end{equation}
which has a rather clear physical interpretation. 

In principle, the source term \( Q_{{\rm c},i} \) receives a contribution from both secondary and primary components, so that formally we can write \( Q_{\rm c} = Q_{\rm c}^{\rm pri} + Q_{\rm c}^{\rm sec} \). 

The secondary source term for a species \( i \), produced by a primary \( j \) and assuming conservation of kinetic energy per nucleon \( E \), is described by:
\begin{equation}\label{eq:qseccocoon}
Q^{\rm sec}_{{\rm c},i}(E) = \frac{\chi_{\rm c}}{\tau_{\rm c}} \frac{\sigma_{j\rightarrow i}}{m_p} N_{{\rm c},j}(E),
\end{equation}
where \(\sigma_{j\rightarrow i}\) is the cross-section of nucleus \( j \) spallating into nucleus \( i \). For cases involving multiple primaries yielding the same secondary, the contributions are summed across all relevant \( j \).

Conversely, the primary component for a nucleus \( i \) is injected according to a spectrum \( q_i(E) \), normalized to the volume \( V_{\rm c} \) and lifetime \( \tau_{\rm life} \) of the cocoon:
\begin{equation}
Q_{{\rm c},i}^{\rm pri} = \frac{q_i(E)}{V_{\rm c} \tau_{\rm life}}.
\end{equation}

Similarly to what we have described inside the cocoon (the hierarchical first leaky box), we can determine the equilibrium density within the Galaxy (the second leaky box) by adapting equations~\eqref{eq:cocoon} and~\eqref{eq:qseccocoon} to the case of the Galaxy. On the other hand, for the Galactic primary source term, we have to account for particles escaping from all cocoons, and we quantify it as follows:
\begin{equation}\label{eq:qgprimary}
Q_{{\rm G},i}^{\rm pri} = \frac{\tau_{\rm life} \mathcal R V_{\rm c}}{V_{\rm G}} \frac{N_{{\rm c},i}}{\tau_c}.
\end{equation}

Here, \(\mathcal R\) denotes the source rate within the Galactic volume \(V_{\rm G} = 2\pi h R_{\rm G}^2\).

To further elucidate the formalism, we apply it to compute the equilibrium ratio between a purely secondary species \( i \) and a purely primary species \( j \) in the Galaxy. 

For a species \( j \) considered as purely primary, the Galactic density is obtained by combining the primary source term in equation~\eqref{eq:qgprimary} with the Galactic transport equation:
\begin{equation}
N_{{\rm G},j} = \left(1+\frac{\chi_{\rm c}}{\chi_j}\right)^{-1} \left(1+\frac{\chi_{\rm G}}{\chi_j}\right)^{-1} \mathcal R \tau_{\rm G} \frac{q_j(E)}{V_{\rm G}}.
\label{eq:primary}
\end{equation}

Conversely, for a secondary species \( i \), we account for two contributions:
a) secondary production in the interstellar medium (ISM):
\begin{equation}
Q_{{\rm G}, i} = \chi_{\rm G} \frac{\sigma_{j\rightarrow i}}{m_p} \frac{N_{{\rm G},j}}{\tau_{\rm G}},
\end{equation}
and b), the cumulative secondary production inside the cocoons:
\begin{equation}
Q_{{\rm c}, i} =  \frac{\tau_{\rm life} \mathcal R V_{\rm c}}{V_{\rm G}}  \frac{N_{{\rm c},i}}{\tau_c}.
\end{equation}

Thus, the secondary equilibrium density is derived by:
\begin{equation}
N_{{\rm G},i} = \left(1+\frac{\chi_{\rm G}}{\chi_i}\right)^{-1} \tau_{\rm G} \left[ Q_{{\rm c}, i} + Q_{{\rm G}, i} \right].
\end{equation}

To simplify the formalism, we are assuming similar critical grammages \( \chi_i \approx \chi_j \), which leads to the widely utilized secondary-to-primary ratio, crucial for deducing the properties of CR transport:
\begin{equation}
\frac{N_{{\rm G},i}}{N_{{\rm G},j}} = \frac{\sigma_{j\rightarrow i}}{m_p}
\left[  \left(1+\frac{\chi_{\rm c}}{\chi_i}\right)^{-1} \chi_{\rm c} + \left(1+\frac{\chi_{\rm G}}{\chi_i}\right)^{-1}  \chi_{\rm G} \right].
\end{equation}
This ratio, which depends on measurable quantities, \( \chi_i \), and grammages, can be fitted against experimental data to infer values for \( \chi_{\rm G} \) and the energy dependence of \( \chi_{\rm c} \).

When addressing secondary unstable nuclei, such as \(^{10}\)Be, the model must also include radioactive decay. By incorporating the decay timescale \(\tau_d=\tau_{d,0}\gamma\), where \(\gamma\) is the Lorentz factor and \(\tau_{d,0}\) is the lifetime at rest, into equation~\eqref{eq:cocoon}, we derive the equilibrium flux of the radioactive isotope within the cocoon as:
\begin{equation}
N_{{\rm c},i} = \left(1+\frac{\chi_{\rm c}}{\chi_i}+\frac{\tau_{\rm c}}{\tau_d}\right)^{-1}\tau_{\rm c} Q_{{\rm c},i}.
\end{equation}
Similarly, the same approach is used to model transport in the Galaxy, with appropriate adjustments from cocoon to Galactic parameters.

It is noteworthy that in scenarios where \(\tau_{\rm c} \ll \tau_d\), the equilibrium density of the radioactive isotope approaches that of its stable counterpart, as anticipated.

By extending this formalism and assuming that decay within the cocoon is negligible (\(\tau_{\rm c} \ll \tau_{\rm d}\)), the ratio between a secondary unstable isotope \( i^* \) and its stable counterpart \( i \), for instance, \(^{9}\)Be, is found to be:
\begin{equation}
\frac{N_{{\rm G},i^*}}{N_{{\rm G},i}} = \frac{\sigma_{j\rightarrow i^*}}{\sigma_{j\rightarrow i}} \frac{\left(1+\frac{\chi_{\rm G}}{\chi_i}\right)}{\left(1+\frac{\chi_{\rm G}}{\chi_i} + \frac{\tau_{\rm G}}{\tau_d}\right)}.
\end{equation}
This equation demonstrates how, making use of the known grammage from the stable secondary-over-primary ratio, the residence timescale \( \tau_{\rm G} \) can be straightforwardly deduced from this observable.

\subsection{Antimatter}
\label{sec:antimatter}

Analyzing the properties of positrons and antiprotons is essential for deciphering the origins and propagation mechanisms of CRs in the standard model of CR transport as well as in the NLBM. These properties complement the insights gained from studying the ratios of secondary to primary nuclei, such as Be/C and B/C. Notably, the NLBM is frequently cited as a natural explanation for the observed flatness in the antiproton-to-proton and positron-to-proton ratios between approximately 60 GV and a few hundred GV, a phenomenon that has posed challenges for standard CR models.

Unlike secondary nuclei, secondary particles such as positrons and antiprotons do not retain the kinetic energy of their primaries. This necessitates accounting for energy redistribution during interactions, thus requiring the use of full differential cross-sections for their production. The secondary source term for antimatter, applicable both within a cocoon and in the Galaxy, is given by:
\begin{equation}\label{eq:sourceantimatter}
Q_{c/G,i}^{\rm sec}(E) = c \int_0^\infty dE^\prime N_{c/G,i}(E^\prime) n_{\rm c/G} \frac{d\sigma_{i,{\rm H}}(E, E^\prime)}{dE},
\end{equation}
where the differential cross-section \( \frac{d\sigma_{i,{\rm H}}(E, E^\prime)}{dE} \) determines the production rates of positrons and antiprotons from interactions between CR species \( i \) and target hydrogen. We consider only hydrogen (H) and helium (He) as the primary species for antimatter production.

%, while we apply a multiplicative fudge factor {\color{red}\( \xi = 1.4 \)} to account for the interstellar medium (ISM) composition \bs{and production by heavier nuclei. Note that this is an optimistic scenario and should be considered as the upper limit of antimatter production that one can expect with the given CR fluxes}. {\bf This is crazy, needs discussion! BS: Added a sentence, that resolves this in my opinion.}

For the calculations of the production rates of positrons and antiprotons we adopt the most recent analytic fits of cross-sections from collider experiments for the most relevant production channels ~\cite{Korsmeier:2018gcy, Orusa:2022pvp}.

For the sole purpose of illustrating the NLBM's predictions, here we consider protons as the only primary particles for antiproton production. Using the source term from equation~\eqref{eq:sourceantimatter}, the Galactic density of \(\bar{p}\) can be calculated as follows:
\begin{equation}\label{eq:NLBM_pbar}
N_{{\rm G},\bar p} = \tau_{\rm G} (Q^{\rm sec}_{{\rm c}, \bar p} + Q^{\rm sec}_{{\rm G}, \bar p}) = 
\frac{1}{m_p} \int_{E_{\bar p}^{\rm th}}^\infty dE^\prime N_{{\rm G}, p}(E^\prime) \left[ \chi_{\rm G} + \chi_c(E^\prime) \right] \frac{d\sigma_{p,\bar{p}}(E, E^\prime)}{dE},
\end{equation}
where \( E_{\bar p}^{\rm th} \) is the minimum energy to produce an antiproton of energy \( E \).%~(see appendix~\ref{app:apth}). 

It is crucial to keep in mind that the mean energy of secondary antiprotons produced in $pp$ collisions is approximately twenty times lower than that of the parent protons. Therefore, if, as demonstrated in subsequent sections, the cocoon grammage \(\chi_c(E)\) decreases rapidly with energy, the secondary production of antiprotons at high enough energies is predominantly determined by transport in the Galaxy rather than in the cocoons~\cite{Cowsik:2016wso}.  

In the case of positrons, modifications similar to those for unstable nuclei are required due to energy losses, that introduce a new timescale, \(\tau_{\rm loss}\), into the model. The flux of positrons in the Galaxy is thus described by:
\begin{equation} \label{eq:NLBM_e+}
N_{{\rm G}, e^+} 
= \frac{1}{m_p} \left( 1+ \frac{\tau_{\rm G}}{\tau_{\rm loss}}\right)^{-1} \int_0^\infty dE^\prime N_{{\rm G}, p}(E^\prime) \left[ \chi_{\rm G} + \chi_c(E^\prime) \right] \frac{d\sigma_{p,e^+}(E, E^\prime)}{dE},
\end{equation}
where energy losses inside cocoons are considered negligible.

Assuming that synchrotron and Inverse Compton scattering on a superposition of interstellar radiation fields are the predominant loss mechanisms, with a photon energy density of \(\sim 0.3 \, \text{eV/cm}^3\) and an average magnetic field of \(1 \, \mu\text{G}\), the timescale for energy losses can be approximated as
\begin{equation}
\tau_{\rm loss} \simeq 1 \, \text{Myr} \left( \frac{E}{\rm TeV} \right)^{-1}.
\end{equation}

This observation stimulated the hyphotesis that the pronounced break in the electron spectrum at 1 TeV~\cite{HESS:2008ibn,DAMPE:2017fbg,CALET:2023emo} might correspond to a transition from a diffusion-dominated to an energy loss-dominated regime at higher energies, potentially establishing \( \tau_{\rm G} \sim 1 \) Myr as a fundamental parameter in various NLBM formulations.

Thus, below 1 TeV, the condition \( \tau_{\rm G} \ll \tau_{\rm loss} \) might help explain the observed flatness in the antiproton to positron ratio, as demonstrated by comparing equation~\eqref{eq:NLBM_pbar} with equation~\eqref{eq:NLBM_e+}. 
Indeed, if energy losses are negligible, the ratio could be primarily influenced by the energy scaling of the cross-section, assumed to be similar across the relevant interactions.

%Due to the minimal cocoon grammage at higher energies, the comparative magnitudes of \(\tau_c\) and \(\tau_l\) do not significantly affect the outcome.

%On the other hand, if energy losses become significant within the Galaxy, such that \(\tau_G \gg \tau_l\), the positron flux approaches zero. Conversely, the standard NLBM often assumes \(\tau_G \ll \tau_l\) to account for the observed flatness in the antiproton to positron ratio. Under these conditions, Equations \ref{eq:NLBM_pbar} and \ref{eq:NLBM_e+} primarily differ in their production cross sections. A flat ratio between antiprotons and positrons can be maintained if these cross sections exhibit similar energy scaling.

%\newpage

%%%%%%%%%% SECTION %%%%%%%%%%
\section{NLBM Comparison with Data}
\label{sec:results}

\subsection{Secondary-over-primary ratios}
\label{sec:sec-to-prim}
\begin{figure}
\centering
\includegraphics[width=0.48\textwidth]{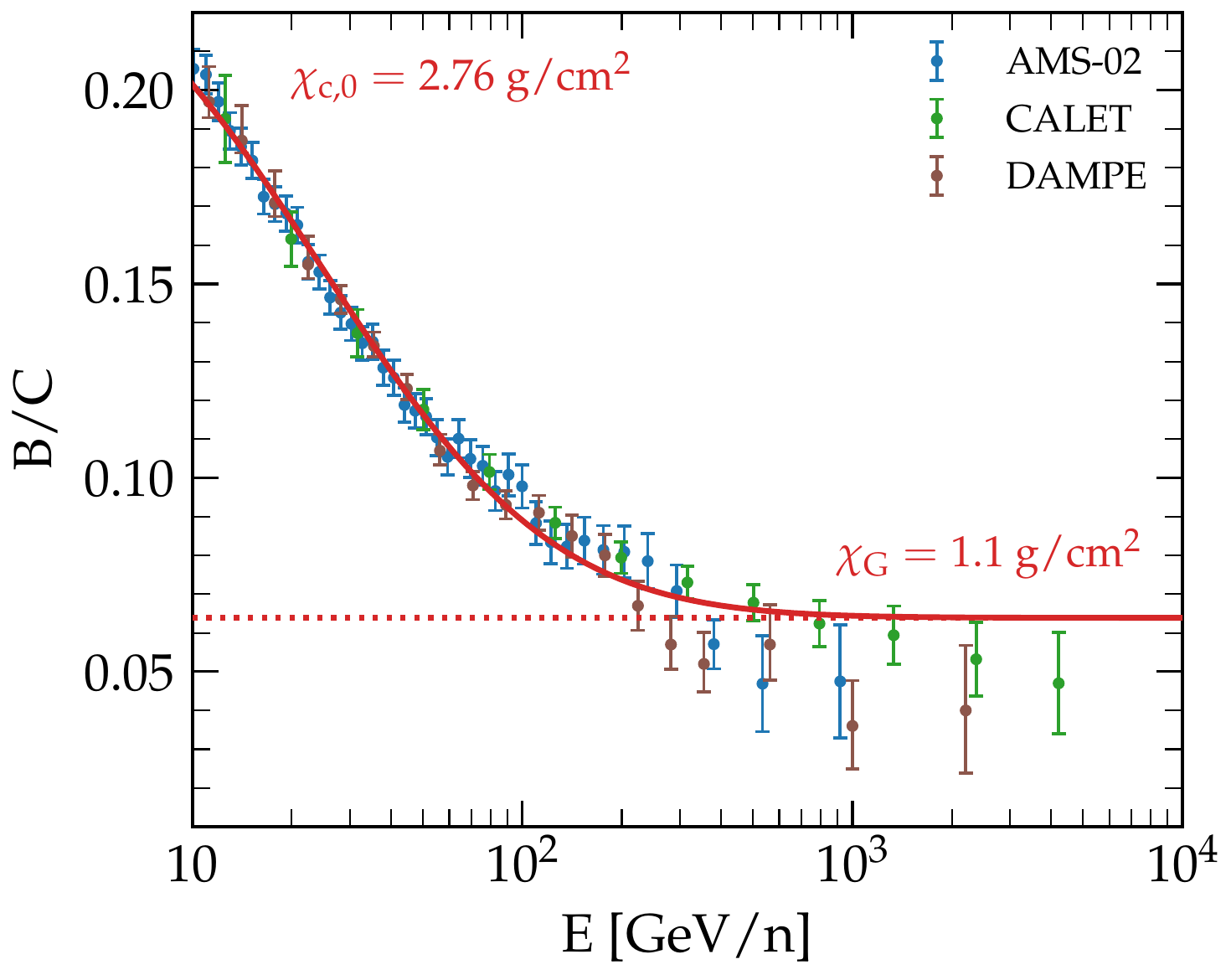}
\includegraphics[width=0.48\textwidth]{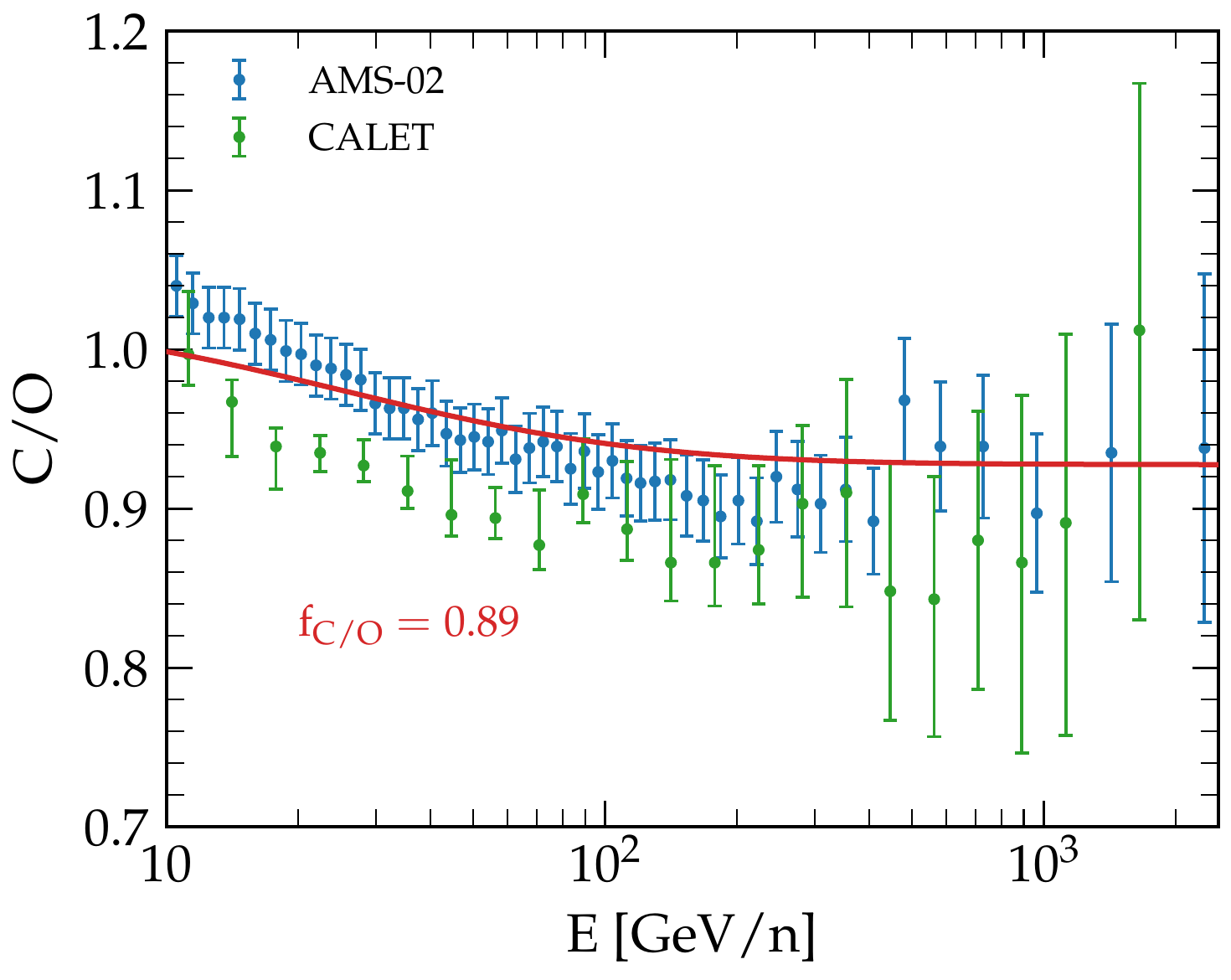}
\caption{Flux ratios of Boron-to-Carbon (B/C, left panel) and Carbon-to-Oxygen (C/O, right panel) as a function of kinetic energy per nucleon, measured by AMS-02~\cite{AMS:2016brs,AMS:2021nhj}, DAMPE~\cite{DAMPE:2022vwu} , and CALET~\cite{CALET:2022dta}. The results are compared with our best-fit model predictions.}
\label{fig:BCO}
\end{figure}

Since the ratios of fluxes of stable secondary nuclei to their primary parent nuclei are well known indicators of the grammage accumulated by CRs, we start our analysis from such ratios, keeping in mind that they are insensitive to changes of slope in the spectra of parent nuclei: all features observed in these ratios result from changes in CR transport. For the sake of keeping the discussion simple and yet quantitatively appropriate, here we restrict our calculations to nuclei lighter than Oxygen and we infer the values of the basic parameters of the NLBM. We cross-checked our results against a model including heavier nuclei.
However, extending the analysis to heavier primaries alters the results by only approximately 10\% and such changes only minimally influence our overall conclusions, while making the calculations and the formalism more cumbersome. Hence, we present the results obtained with the simplified formalism for the sake of clarity.

Following the methodology of Cowsik et al.~\cite{Cowsik:2016wso}, we model the energy dependence of the cocoon grammage as:
\begin{equation}
\chi_c(E) = \chi_0 \left( \frac{E}{10~{\rm GeV}}\right)^{-\zeta \ln \left( \frac{E}{{\rm GeV}} \right)}\,.
\label{eq:Xcoc}
\end{equation}

Using a power law scaling for the diffusion coefficient, as one would expect from standard resonant scattering in a power law spectrum of perturbations, does not affect our findings qualitatively. Thus, in order to retain consistency with established literature, we adopt the functional form in equation~\eqref{eq:Xcoc} for the grammage in the cocoon regions.

We also focus on energies above \( \gtrsim 10 \, \text{GeV/n} \), where ionization energy loss and Solar modulation have minimal effects on the nuclear ratios. In this regime, secondary nuclei produced by spallation retain nearly the same kinetic energy per nucleon as their progenitor nuclei.

We consider Boron as a purely secondary species, with Carbon and Oxygen as primaries. The Boron-to-Carbon (B/C) and Carbon-to-Oxygen (C/O) ratios are pivotal for constraining the model parameters. Making use of the results in section~\ref{sec:model}, we write these ratios as follows:
\begin{equation} \label{eq:NLBM_BC}
\frac{\rm B}{\rm C} = \left(\frac{\sigma_{\rm C \rightarrow B}}{m_p} + \frac{\rm O}{\rm C} \frac{\sigma_{\rm O \rightarrow B}}{m_p} \right) 
\left[ \chi_{\rm G} \left( 1 + \frac{\chi_{\rm G}}{\chi_{\rm B}} \right)^{-1} + \chi_c \left( 1 + \frac{\chi_{\rm c}}{ \chi_{\rm B}} \right)^{-1} \right] \, , 
\end{equation}
and 
\begin{equation} \label{eq:NLBM_CO}
\frac{\rm C}{\rm O} = f_{\rm CO} + \frac{\sigma_{\rm O \rightarrow C}}{m_p} 
\left[ \chi_{\rm G} \left( 1 + \frac{\chi_{\rm G}}{\chi_{\rm C}} \right)^{-1} + \chi_{\rm c} \left( 1 + \frac{\chi_{\rm c}}{ \chi_{\rm C}} \right)^{-1} \right]  \,.
\end{equation}

The latter equation also includes the secondary production of Carbon from the spallation of Oxygen nuclei. For the sole purpose of simplifying the formalism, while introducing negligibly small approximations, we assume \( \chi_{\rm B} \simeq \chi_{\rm C} \) in equation~\eqref{eq:NLBM_BC} and \( \chi_{\rm C} \simeq \chi_{\rm O} \) in equation~\eqref{eq:NLBM_CO}. This assumption introduces an uncertainty of approximately 10\% in the critical grammages.
The critical grammage is calculated adopting the simple geometrical approximation for the inelastic cross-section~\cite{Letaw1983ApJS}:
\begin{equation}\label{eq:criticalgrammage}
\chi_i \sim \frac{m_p}{45 A_i^{2/3} {\rm mb}} \sim 8 \left( \frac{A_i}{10} \right)^{-2/3}~{\rm g}~{\rm cm}^{-2}
\end{equation}

Concerning the production cross-sections we adopt the model described in~\cite{Evoli:2019wwu} as our baseline and we report the numerical values in Table~\ref{tab:cross_sections} to facilitate reproducibility of our results. The cross-sections are assumed to be energy-independent within the considered energy range, in line with the common practice in the field~\cite{Genolini:2018ekk}.

\begin{table}[ht]
\caption{\label{tab:cross_sections} Relevant Cross-Sections in the Model computed following~\cite{Evoli:2019wwu}.}
\begin{ruledtabular}
\begin{tabular}{l|ccccccc}
Channel & C$\rightarrow$O & O$\rightarrow$B & C$\rightarrow$B & O$\rightarrow$$^{10}$Be & O$\rightarrow$$^{9}$Be & C$\rightarrow$$^{10}$Be  & C$\rightarrow$$^{9}$Be \\
Cross-Section (mb) & 60 & 37 & 71 & 4 & 3 & 4 & 7 \\
\end{tabular}
\end{ruledtabular}
\end{table}

We utilize measurements from AMS-02~\cite{AMS:2016brs,AMS:2021nhj}, CALET~\cite{CALET:2022dta}, and DAMPE~\cite{DAMPE:2022vwu}. For each data point, we calculate error bars by summing the statistical and systematic uncertainties in quadrature, as reported by these collaborations. In doing so, we assume that the systematic errors are entirely uncorrelated. The minimization procedure employs the MINUIT algorithm~\footnote{\url{https://github.com/jpivarski/pyminuit}}. We perform fits by minimizing the combined \( \chi^2 \) for the ratios B/C and C/O, over a parameter space that includes \( \chi_{\rm G} \), \( \chi_{{\rm c},0} \), \( \zeta \), and the relative normalization \( f_{\rm CO} \) of the primary injection spectra. 

Our best-fit scenario is illustrated in Fig.~\ref{fig:BCO} and corresponds to the following parameter values: \( \chi_{\rm G} = 1.1 \, \text{g/cm}^2 \), \( \chi_{{\rm c},0} = 2.76 \, \text{g/cm}^2 \), \( \zeta = 0.18 \), and \( f_{\rm CO} = 0.89 \).

The left panel of Fig.~\ref{fig:BCO} presents our results for the B/C ratio, while the right panel offers a comparison for the C/O ratio, as predicted by equations~\ref{eq:NLBM_BC} and \ref{eq:NLBM_CO}. Both species are assumed to be injected with the same spectral slope. The notable low-energy trend in the C/O ratio, characterized by a deviation from a constant ratio, is effectively explained by the spallation of Oxygen into Carbon nuclei.

We confirm that the NLBM provides an accurate description of the available B/C data, however implying Galactic residence times significantly shorter than those posited by the traditional halo model. Indeed, consistent with prior studies, and assuming an average Galactic density of \( n_{\rm G} \sim 0.5 \, \text{cm}^{-3} \), our best-fit scenario indicates a Galactic residence time \( \tau_{\rm G} \sim \) Myr. This timescale is several orders of magnitude shorter than the propagation time estimated in the halo model for \( \sim 10 \) GeV particles within a low-density halo with \( H \simeq 4 \, \text{kpc} \).

However, as discussed in Section~\ref{sec:model}, the significantly shorter cosmic-ray residence time in the Galaxy matches the energy-loss timescale for TeV electrons through inverse Compton scattering on the galactic radiation fields and synchrotron radiation in a \( \sim 1 \mu \text{G} \) magnetic field. This observation motivated the identification of the spectral break observed in the \( e^- + e^+ \) spectrum at \( E \sim 1 \, \text{TeV} \) as evidence supporting the NLBM, particularly due to the onset of energy losses. 
While our findings confirm that this scenario is not in conflict with models that posit secondary nuclei production through nuclear fragmentation in the cocoons, subsequent sections will discuss how this model is incompatible with observations of secondary antimatter and radioactive Beryllium isotopes.

Finally, it is important to emphasize that the hardening observed at a few hundred GeV/n in the B/C ratio is modeled in the NLB as a transition from secondary nuclei produced within the cocoons at lower energies to those generated during propagation through the Galactic disk. Thus, this feature cannot be related to the similar break observed in the spectrum of primary species. In the context of the NLBM, any such similarity would be merely coincidental.

In contrast, the halo model typically attributes this feature to a change in CR propagation properties~\cite{Genolini:2017dfb}, possibly linked to a transition in the origin of turbulence~\cite{Blasi:2012yr} or to a spatial dependence of the diffusion coefficient~\cite{Tomassetti:2012ga}. In this scenario, the model predicts that the same break should manifest equally in both the secondary-over-primary ratio and the primary spectra, as observed (see the following section). 

\subsection{Spectra of primary protons and Helium nuclei}\label{sec:p+He}

\begin{figure}
\centering
\includegraphics[width=0.48\textwidth]{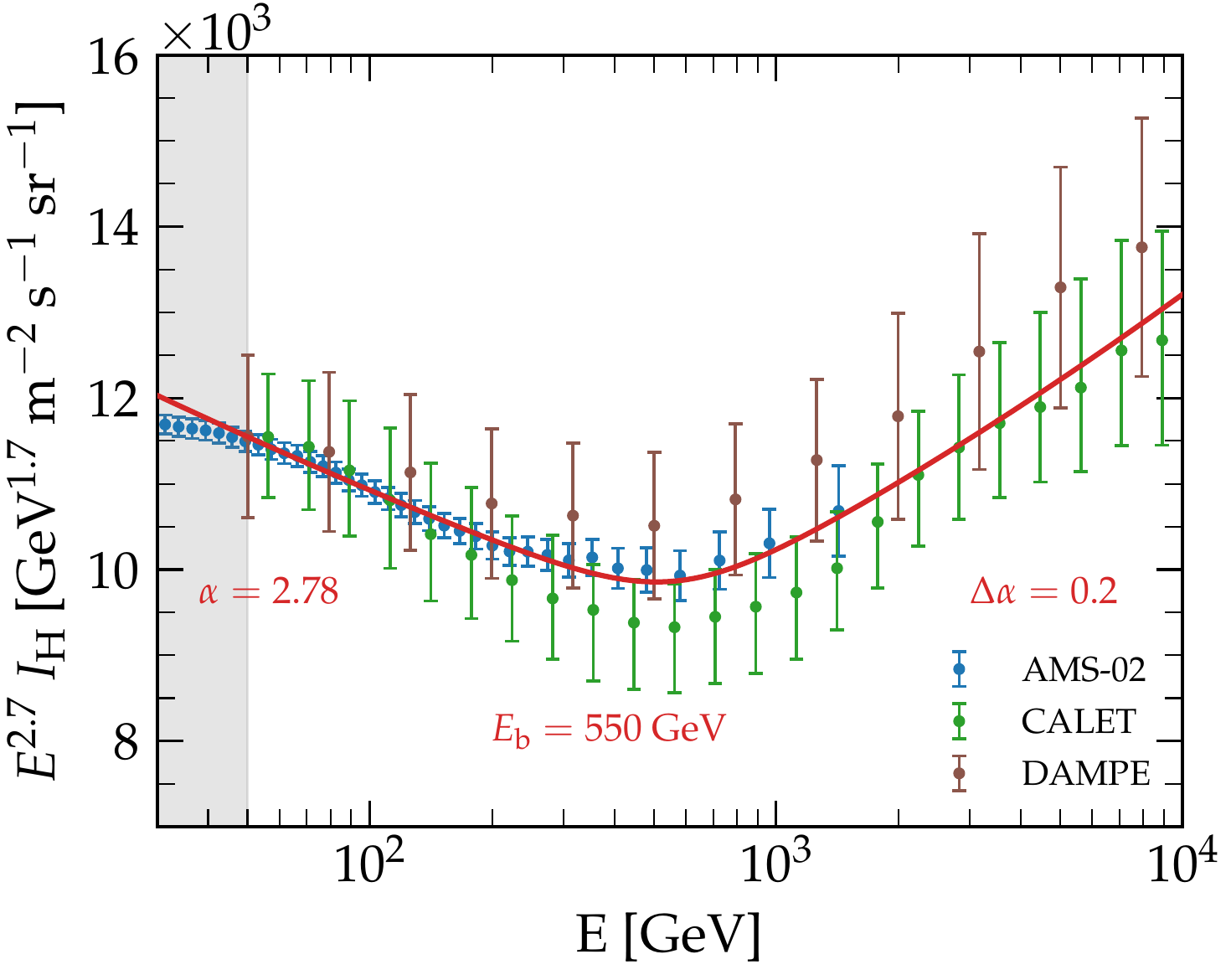}
\includegraphics[width=0.48\textwidth]{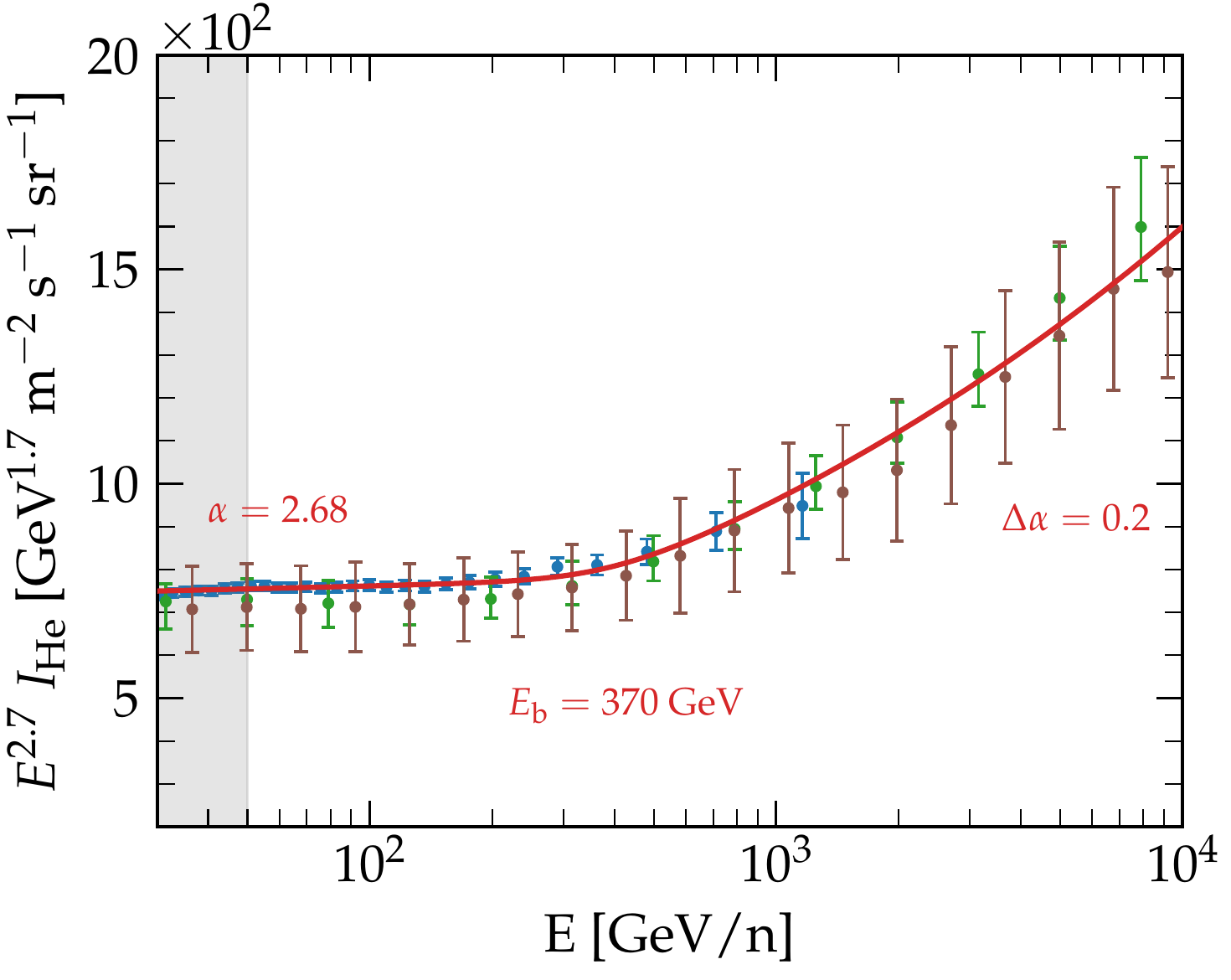}
\caption{Flux of protons (left panel) and helium nuclei (right panel) as a function of kinetic energy per nucleon. The data points encompass measurements from the AMS-02~\cite{AMS:2021nhj}, DAMPE~\cite{DAMPE:2019gys,DAMPE:2023pjt} and CALET~\cite{CALET:2019bmh,CALET:2023nif} experiments. Solid lines show our best-fit model predictions as described in Sec.~\ref{sec:p+He}. The shaded region delineates the energy range excluded from our fitting process.}
\label{fig:pHe}
\end{figure}

In the version of the NLBM that is commonly applied to describe the secondary/primary ratios and the production of antimatter, primary nuclei are assumed to be injected with a given power law. For protons and helium nuclei, where energy losses have little effect, the NLBM predicts that the observed spectra are the same as the ones injected at the sources. Accurate modeling of proton and helium spectra is critical for at least two reasons: first, it provides insights into the acceleration mechanisms that operate in supernova remnants (SNRs) and potentially other sources; second, it allows for a robust determination of the source term for antimatter production (e.g., positrons and antiprotons), since protons and helium nuclei are the main progenitors of these secondary particles. Here we test the predictions of the NLBM against existing data. 

First we notice that observations indicate a change in the proton spectral slope around \( E \sim 500 \,\text{GeV} \), first noted by PAMELA~\cite{PAMELA:2011mvy}, confirmed with higher precision by AMS-02~\cite{AMS:2021nhj} and now observed by DAMPE~\cite{DAMPE:2019gys,DAMPE:2023pjt} and CALET~\cite{CALET:2019bmh,CALET:2023nif} as well. In the context of the NLBM, as discussed above, the equilibrium spectrum of protons and He nuclei must reproduce the source spectrum, hence we are forced to conclude that the spectral break must reflect a feature in the injection process itself.

Hence, we parametrize the injection spectrum of a primary species \( i \) by:
\begin{equation}
q_i(E) \;=\; q_{0,i} \left( \frac{E}{E_0} \right)^{-\alpha_i} 
\left[ 1 + \left( \frac{E}{E_{b,i}} \right)^{\frac{\Delta \alpha}{s}} \right]^s,
\end{equation}
where \( E_{b,i} \) is the break energy, \( \alpha_i \) is the spectral index, and \( s \) is the smoothness of the transition, fixed at \( s = 0.05 \). 

The normalization $q_{0,i}$ at energy $E_0=1\,$GeV is calculated such that:
\begin{equation}
\int_{E_0}^\infty \! dE \, A E \, q_p(E) \;=\; \xi E_{\rm CR}
\;\;\rightarrow\;\;
q_{0,i} \;\simeq\; \frac{\xi_i \, E_{\rm CR} \, (\alpha_i - 2)}{A \, E_0^2},
\label{eq:normalization}
\end{equation}
where \( \xi_i \) is the efficiency of CR acceleration and \( E_{\rm CR} = 10^{51} \,\text{erg} \) is the total energy of a supernova (SN) explosion (excluding neutrinos).

The propagation of proton and helium follows the primary-component description presented in the previous section assuming \( \tau_{\rm G} \sim 1 \)~Myr. In particular, we neglect inelastic proton interactions, since the corresponding timescale is much longer than any other relevant timescale in this scenario. For helium, we adopt a critical grammage \(\chi_{\rm He} = 15\,\text{g}\,\text{cm}^{-2}\) (see equation~\ref{eq:criticalgrammage}). 

Using the parameterization of Eq.~\ref{eq:primary}, we obtain the following best-fit parameters for protons: \(\xi \approx 4\%\), \(\alpha = 2.78\), \(E_b = 550 \,\text{GeV}\), and \(\Delta \alpha = 0.2\). For helium, the fit yields \(\xi \approx 0.3 \% \), \( \alpha = 2.68 \), \( E_b = 370 \, \text{GeV} \), and \( \Delta \alpha = 0.2 \) (see Fig.~\ref{fig:pHe}).

We first note that the overall acceleration efficiency—of the order of a few percent of the SN kinetic energy—is consistent with previous estimates (e.g.,~\cite{Blasi:2013rva,Gabici:2019jvz}). From Eq.~\ref{eq:primary}, it is clear that \(\xi\) remains degenerate with the ratio of confinement time to confinement volume, which in the NLBM is effectively the size of the disk. Although both the numerator and denominator in this ratio are smaller in the NLBM compared to conventional diffusion-halo models, the resulting efficiency stays at a similarly plausible level. One caveat to this conclusion is that the slope $\alpha_i$ at the source is much larger than in the halo model.
In this case, one should not neglect the non-relativistic part of the spectrum and start the integration in Eq.~\ref{eq:normalization} at the injection energy and not $E_0$. However we checked that this leads to an underestimate of the actual efficiency by less than a factor $\sim 2$, so that the value of $\xi$ remains plausible even in a more refined estimate. 

An immediate consequence of the energy-independent propagation in the NLB scenario is that the observed spectral shape must directly mirror the injection spectrum. Inelastic losses for protons at these energies are minimal and do not significantly change the equilibrium spectrum. It follows that the NLBM requires that {\it a)} the source spectrum is a broken power law, and {\it b)} the overall slope is very steep, unlike the one expected from diffusive shock acceleration (DSA) in SNRs or other sources. Both these aspects are at odds with current knowledge of how particle acceleration works. 

Finally, the NLBM shares a problematic aspect with the standard halo model in that the required source spectra of protons and He nuclei are different. It has been speculated that this may have something to do with the efficiency of injection of nuclei as a function of the shock Mach number during the Sedov phase of a SNR~\cite{Malkov:2011gb}. Such model would however be hardly compatible with the overall steep spectrum required by the NLBM. 

One last point that we want to stress and is of the highest importance is that, as noted in Section~\ref{sec:sec-to-prim}, the break in the proton spectrum at \( E \sim 500 \,\text{GeV} \) appears at the same energy as the break observed in secondary-to-primary ratios, such as B/C. In the halo model, a shared propagation-related origin explains both features simultaneously. In the NLB scenario, however, this coincidence should be treated as fortuitous.

\subsection{Unstable Nuclei: The Case of $^{10}$Be}

\begin{figure}
\centering
\includegraphics[width=0.48\textwidth]{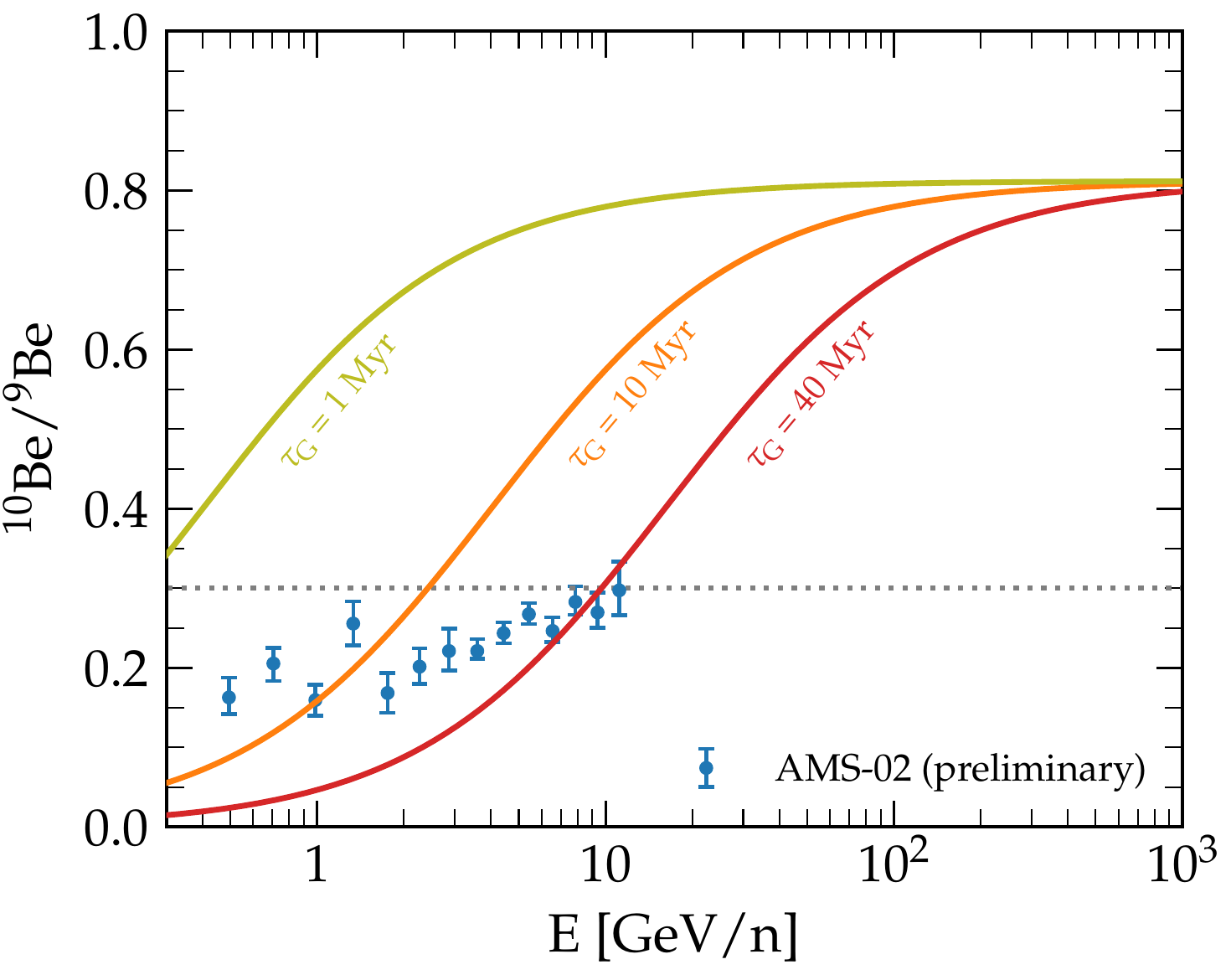}
\caption{Flux ratios of beryllium isotopes (\(^{10}\mathrm{Be}/^{9}\mathrm{Be}\)) as a function of kinetic energy per nucleon. The plot displays preliminary measurements from the AMS-02 experiment. Predictions from the NLB model are overlaid for comparison, illustrating different galactic residence timescales (\(\tau_{\mathrm{G}}\)). The curves, from top to bottom, correspond to \(\tau_{\mathrm{G}} = 40\), \(10\), and \(1\) Myr, respectively.}
\label{fig:Be}
\end{figure}

The flux of the unstable isotope of beryllium (\(^{10}\)Be) has long been recognized as a crucial diagnostic of the nature of CR propagation through the Galaxy. Unlike stable isotopes such as \(^{9}\)Be and \(^{7}\)Be, \(^{10}\)Be has a relatively short half-life of \( T_{1/2} \simeq 1.387 \pm 0.012 \) million years, which is comparable to the typical residence time of CRs in the Galactic disk. This correspondence allows the radioactive decay of \(^{10}\)Be to serve as a natural clock, placing direct constraints on the time CRs spend in the Galaxy before escaping or undergoing significant energy losses.

In fact, by comparing the spectral shape of \(^{10}\)Be with that of stable beryllium isotopes, such as \(^{9}\)Be and \(^{7}\)Be, one can infer the degree of radioactive decay and, consequently, estimate the Galactic residence time of CRs. This measurement is fundamental for testing and differentiating between competing CR propagation models, including the NLBM and standard diffusion models.

Despite the importance of these measurements, they have received relatively limited attention in the context of the NLBM development. One reason for this may be the experimental challenges involved in measuring the flux of individual isotopes, particularly at high energies where the CR flux becomes increasingly rare and isotopic separation becomes more difficult. To date, most beryllium isotope measurements have been restricted to low-energy regimes, with significant statistical and systematic uncertainties, thereby limiting their ability to robustly distinguish between propagation models.

The highest-energy data available for \(^{10}\)Be come from the ISOMAX balloon experiment, which measured the \(^{10}\)Be/\(^{9}\)Be ratio as \( 0.317 \pm 0.109 \) in the energy range \( \sim 1-2 \, \text{GeV/n} \)~\cite{Hams:2004rz}. While the ISOMAX data marked an important step forward, the large uncertainties and broad energy bins reduce the constraining power of these measurements.

Looking ahead, the CR community eagerly awaits more precise measurements from experiments such as AMS-02 and the HELIX balloon mission~\cite{HELIX:2023twg}, both of which are expected to provide high-resolution isotopically separated spectra of beryllium up to \( \gtrsim 10 \, \text{GeV/n} \). These forthcoming datasets are anticipated to impose much tighter constraints on CR propagation models, especially with regard to the Galactic residence time, and are expected to resolve many of the current uncertainties in propagation modeling.

In the meanwhile, we decided to use the preliminary results from AMS-02~\cite{Dimiccoli:2023dbx}, in order to illustrate a preview of what to expect when better data, extending to higher energies, will become available. These preliminary data focus on beryllium isotopic ratios in the energy range 0.5–10 GeV/n, with the highest energy points suggesting a \(^{10}\)Be/\(^{9}\)Be ratio of approximately \( \sim 0.3 \) at around 10 GeV/n (see figure~\ref{fig:Be}).

At the energies of interest, it is reasonable to assume that all beryllium isotopes form predominantly via spallation of heavier nuclei—primarily carbon and oxygen—when they collide with gas in the cocoons, while the equilibrium spectra are shaped by propagation in the Galaxy. Under these conditions, the ratio of an unstable to a stable isotope can be expressed as
\begin{equation}\label{eq:beratio}
\frac{^{10}\mathrm{Be}}{^{9}\mathrm{Be}} \;=\; 
\frac{\sigma_{\mathrm{O}\to 10} + f_{\mathrm{CO}}\,\sigma_{\mathrm{C}\to 10}}{\sigma_{\mathrm{O}\to 9} + f_{\mathrm{CO}}\,\sigma_{\mathrm{C}\to 9}}
\;\frac{1 + \tfrac{\chi_{\mathrm{G}}}{\chi_{9}}}{1 + \tfrac{\chi_{\mathrm{G}}}{\chi_{10}} + \tfrac{\tau_{\mathrm{G}}}{\gamma(E)\,\tau_{d,0}}}
\;\simeq\; 
0.8 \;\frac{1 + \tfrac{\chi_{\mathrm{G}}}{\chi_{9}}}{1 + \tfrac{\chi_{\mathrm{G}}}{\chi_{10}} + \tfrac{\tau_{\mathrm{G}}}{\tau_{d}}}~,
\end{equation}
where \(\tau_{d,0} \sim 2~\mathrm{Myr}\) is the decay lifetime at rest, \(\gamma\) is the Lorentz factor. 

%To reproduce the observed \(\,^{10}\mathrm{Be}/\,^{9}\mathrm{Be}\) ratio of approximately 0.3, one requires a Galactic residence time \(\tau_{\mathrm{G}} \gtrsim 40~\mathrm{Myr}\) at these energies, given that \(\tau_d \simeq 20~\mathrm{Myr}\). If future measurements confirm this value, it would strongly suggest a much longer Galactic residence time than is typically expected in the NLBM.

Figure~\ref{fig:Be} illustrates the predicted \(\,^{10}\mathrm{Be}/\,^{9}\mathrm{Be}\) ratio based on equation~\eqref{eq:beratio} for different values of the Galactic CR residence time \(\tau_{\mathrm{G}}\). At around 10~GeV/n—where our cross-section calculations are most robust—the best agreement occurs for \(\tau_{\mathrm{G}} \gtrsim 10~\mathrm{Myr}\). Conversely, adopting \(\tau_{\mathrm{G}} \sim 1~\mathrm{Myr}\) overestimates the measured ratio by more than a factor of two.
If future measurements confirm this value, it would strongly suggest a much longer Galactic residence time than is typically expected in the NLBM~\cite{Lipari:2014zna}.
At lower energies, the model fails to track the data trend accurately, likely because it assumes energy-independent cross sections. A more sophisticated cross-section treatment is beyond the scope of this work and is further complicated by existing experimental uncertainties. Nonetheless, this approach will become increasingly relevant as upcoming measurements from AMS-02 and the HELIX mission provide more precise constraints on Galactic isotopic ratios.

A complementary approach to this analysis was proposed by Webber and Soutoul  in 1998~\cite{Webber:1998ex}, who suggested the use of elemental ratios (e.g., Be/B, Al/Mg) that incorporate both the decayed fraction of the unstable isotope and the enhanced contribution from decay products. These elemental ratios have already been measured by AMS-02 up to several hundred GeV/n, providing critical constraints on CR propagation models. In this context, in~\cite{Evoli:2019iih} we demonstrated how these measurements span the energy range in which \(^{10}\)Be transitions from being largely decayed at lower energies to relatively stable at higher energies, consistently yielding estimates of the residence time  \( \tau_{\mathrm{res}} \) that are well in excess of 20 Myr.

\subsection{Anti-matter fluxes in Galactic CRs}

\begin{figure}
\centering
\includegraphics[width=0.48\textwidth]{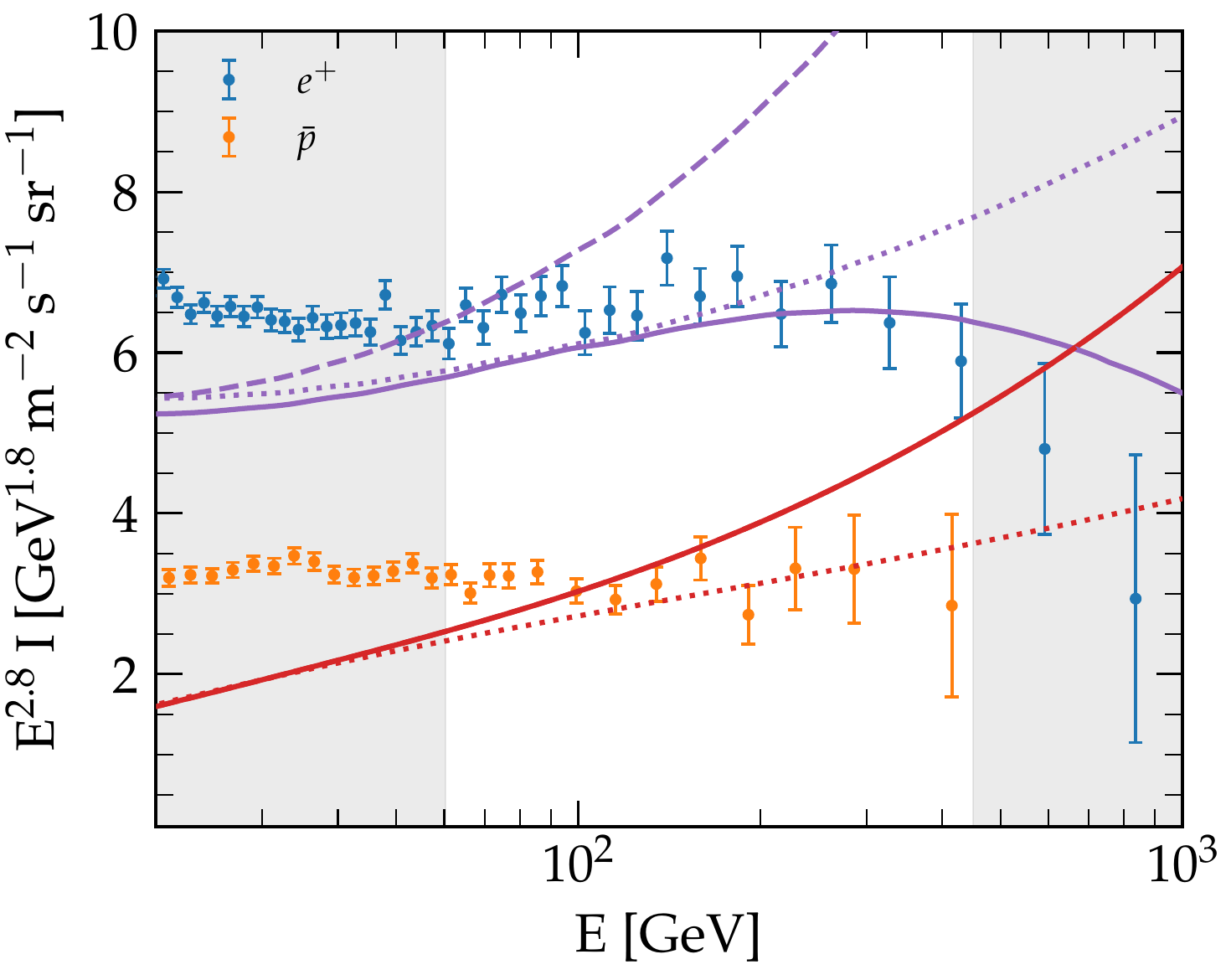}
\includegraphics[width=0.48\textwidth]{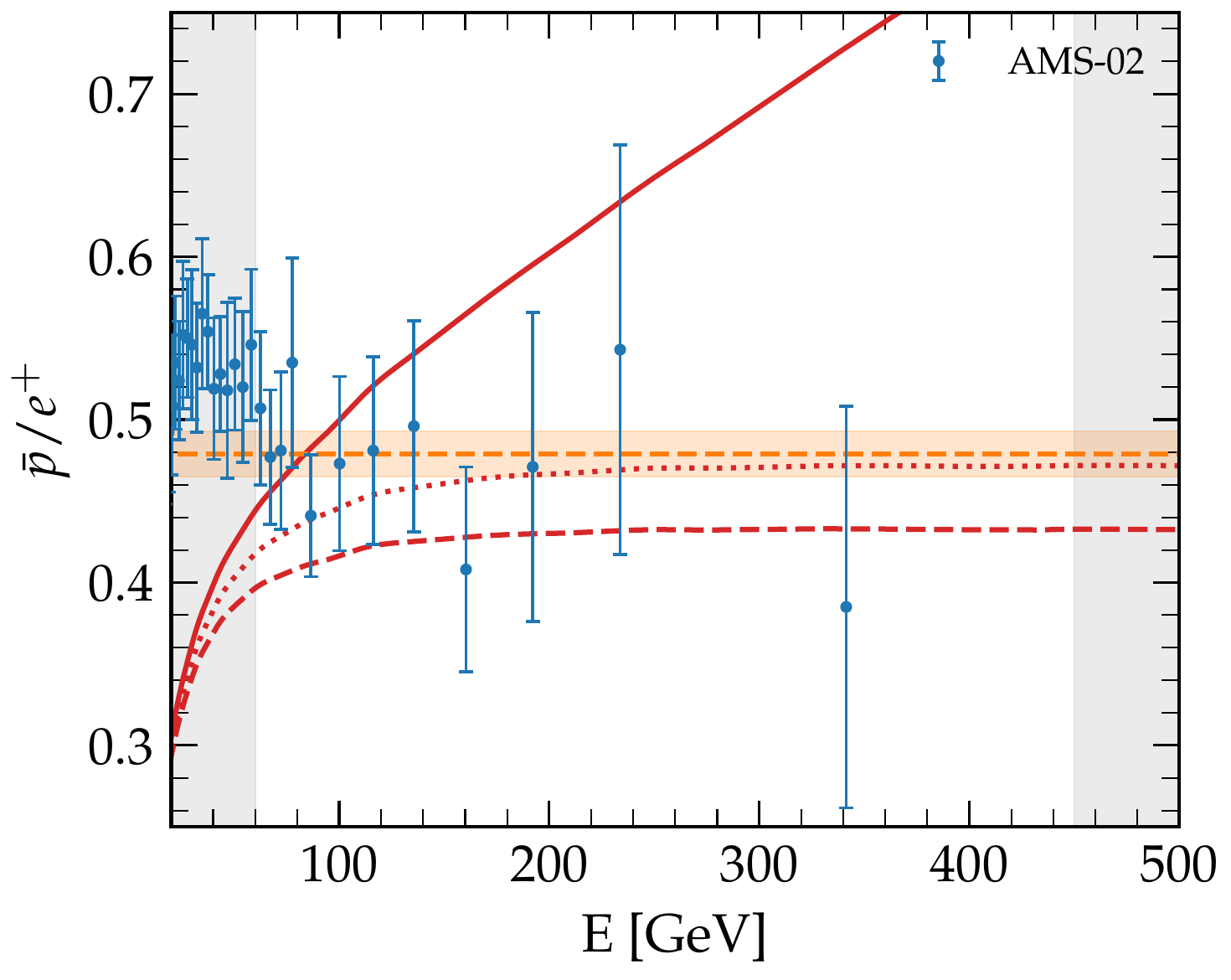}
\caption{Fluxes of antiprotons (\(\bar{p}\)) and positrons (\(e^+\)) (left panel), along with their flux ratio (right panel) as functions of energy, compared with the predictions of the NLBM. The model representations are as follows: dotted lines for predictions without a break in the primary spectrum and without energy losses in the positron model, dashed lines including only the break in the primary spectrum, and solid lines including both the break and energy losses in the positron model. The model fluxes have been rescaled by a factor of \( \eta = 2 \), as detailed in the text. The orange dashed line in the right panel denotes the best-fit constant value \(K = 0.479 \pm 0.014\), applicable over the energy range \(60.3 \leq E \leq 450 \, \text{GeV}\), as reported by AMS-02~\cite{AMS:2016oqu}. The gray shaded areas highlight energies outside the validity range of the constant fit, depicted in both panels.}
\label{fig:pbar_e+}
\end{figure}

In this section, we focus on the production and propagation of antimatter, specifically positrons and antiprotons. The NLBM is frequently invoked as a \emph{natural} solution to the nearly flat antiproton-to-positron ratio observed between \(\sim 60\)\,GeV and a few hundred GeV~\cite{Cowsik:2016wso}. Following Section~\ref{sec:antimatter}, we calculate the antiproton and positron fluxes arising from primary proton and helium fluxes in the NLBM, with parameters tuned to match the previously discussed CR ratios.

Figure~\ref{fig:pbar_e+} compares our predictions for the antiproton and positron fluxes (left panel), as well as their ratio (right panel), to recent AMS-02 data. The right panel shows that, if one neglects energy losses for positrons, the \(\bar{p}/e^+\) ratio remains almost perfectly flat above \(\sim 80\)\,GV, albeit outside the range found by AMS-02 when fitting a constant value of \(K = 0.479 \pm 0.014\) in the range \(60.3 \le E \le 450\,\text{GeV}\)~\cite{AMS:2016oqu}. However, this picture changes dramatically once synchrotron and inverse-Compton energy losses for positrons are taken into account even with $\tau_G\sim 1\,$Myr. Around \(100\)\,GV, these processes introduce substantial deviations, spoiling the flatness that is often cited as an element in support of the NLBM.

A further issue arises when comparing the predicted absolute fluxes of \(\bar{p}\) and \(e^+\) to data (left panel of Fig.~\ref{fig:pbar_e+}). Our best-fit model underestimates both fluxes by about a factor of two (the fluxes shown in Fig.~\ref{fig:pbar_e+} are already multiplied by a factor 2). We recall here that neglecting the production of antimatter by nuclei heavier than oxygen only introduces a \(\sim 10\%\) level deficit~\cite{Korsmeier:2018gcy}. Hence, the reason for this deficit is to be searched for elsewhere. We will go back to a discussion of this point below. 

However, aside from the normalization of the flux of secondary antimatter, both the antiproton and positron spectra turn out to be appreciably harder than the observed spectra at low energies, namely below 100 GeV. This poses a serious challenge to the NLBM, in that there is no obvious physical mechanism to alleviate this tension.

The hardening in the spectra of both antiprotons and positrons at energies above \(50\)\,GV is a direct consequence of the break in the primary proton and helium spectra, the main progenitors of secondary antiprotons and positrons. This implies that, in order to establish consistency of the NLBM with the whole set of data, including the spectra of H and He, one is forced to accept harder \(\bar{p}\) and \(e^+\) spectra, at odds with earlier results, in which the spectral breaks of H and He were not accounted for~\cite{Cowsik:2022eik}. 
Including energy losses for positrons improves the agreement of the high-energy positron flux with data, but it breaks the near-constant behavior of the \(\bar{p}/e^+\) ratio. Consequently, once tuned to CR nuclei data, the NLBM struggles to match both the absolute normalization and energy dependence of the measured antiproton and positron fluxes.

\begin{figure}
\centering
\includegraphics[width=0.46\textwidth]{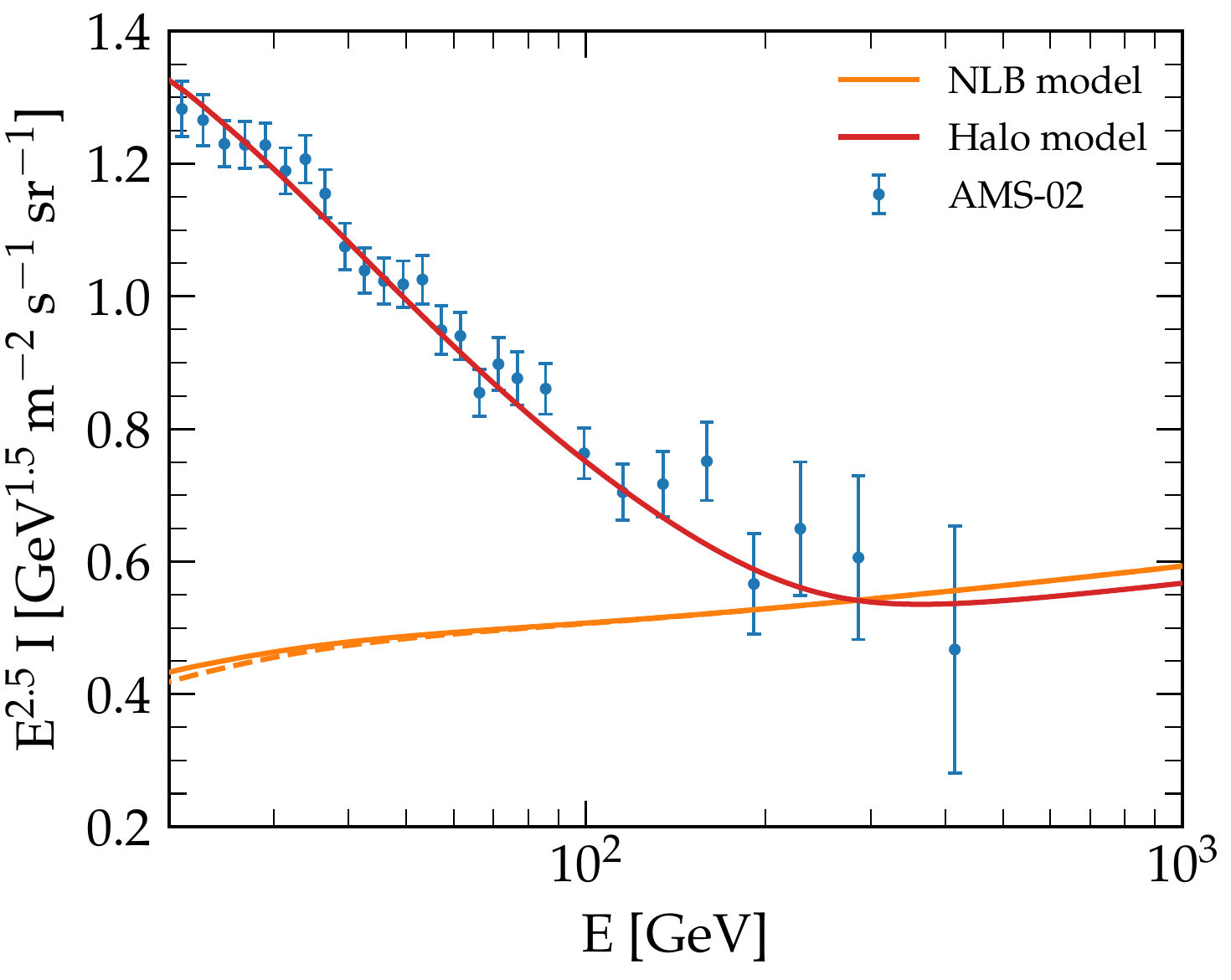}
\caption{Predicted flux of antiprotons in the NLBM and in the halo model compared to the data. A dashed line shows the \emph{cocoon} contribution in the NLB model.}
\label{fig:aphm}
\end{figure}

As discussed above, the NLBM leads to a deficit in the absolute fluxes of positrons and antiprotons, by about a factor of 2. This is an intrinsic feature of the NLBM due to, as we discuss below, the requirement that the escape time from the Galaxy, $\tau_G$, is independent of energy. In order to make this point, we compare the predicted spectrum of antiprotons in the NLBM with that of the standard halo model or a simple leaky box model with an energy dependent Galactic confinement time under the same assumptions.

%{\color{red}Comment on this plot}. \bs{see below}
%\subsection{Normalization of Antimatter Fluxes}
%\bs{The exact reason why the antimatter fluxes of the NLBM fall a factor of a few below the measurements lies in the energy independent $\tau_G$ in Eq.~\ref{eq:NLBM_pbar}. To verify that it is a consequence of the model itself and not our simplifying assumptions, it is instructive to compare the NLBM to the halo model or an equivalent simple leaky box model with an energy dependent Galactic time scale under the same assumptions.

Within the halo model, the B/C ratio can be calculated using Eq.~\ref{eq:NLBM_BC} assuming $\chi_c=0$ and an energy-dependent Galactic residence time, $\tau_{\rm H}(E)$, and grammage, $\chi_H(E)$. It is easy to see that if inelastic losses are negligible, the predicted B/C in the halo model is equivalent to that of the NLBM with roughly the same total grammage $\chi_H=\chi_G+\chi_c$. For antimatter, the equivalent of Eq.~\ref{eq:NLBM_pbar} in the halo model reads
\begin{equation}\label{eq:halo_model}
N_{{\rm H},\bar p} = \tau_{\rm H}(E) Q^{\rm sec}_{{\rm H}, \bar p} = 
\frac{1}{m_p} \chi_H(E) \int_{E_{\bar p}^{\rm th}}^\infty dE^\prime N_{{\rm H}, p}(E^\prime)  \frac{d\sigma_{p,\bar{p}}(E, E^\prime)}{dE}.
\end{equation}

For a given antiproton energy $E$, the prediction of the halo model compared with the NLBM model differs by an energy-dependent factor $\chi_{\rm H}(E)/\chi_{\rm G}$ (assuming for obvious reasons that $\chi_{\rm c}(E^\prime) \ll \chi_{\rm G}$ in Eq.~\eqref{eq:NLBM_pbar}, as in the NLBM antiprotons are mainly produced by protons outside the cocoons).

We illustrate this in Fig.~\ref{fig:aphm}, where we show the predictions for the antiproton flux of the NLBM and the halo model, both tuned to reproduce B/C, compared to the data. Hence, we conclude that the discrepancy found above is intrinsic to the model and not due to any of our simplifying assumptions.
Our finding is compatible with the estimates in~\cite{Diesing:2020jtm}, which suggest that B/C and antiprotons and positrons are incompatible in models like the NLBM based on the average gas density needed to explain the measurements.

\section{Cross sections and antiproton/proton ratio}
\label{sec:xsec}

\begin{figure}
\centering
\includegraphics[width=0.46\textwidth]{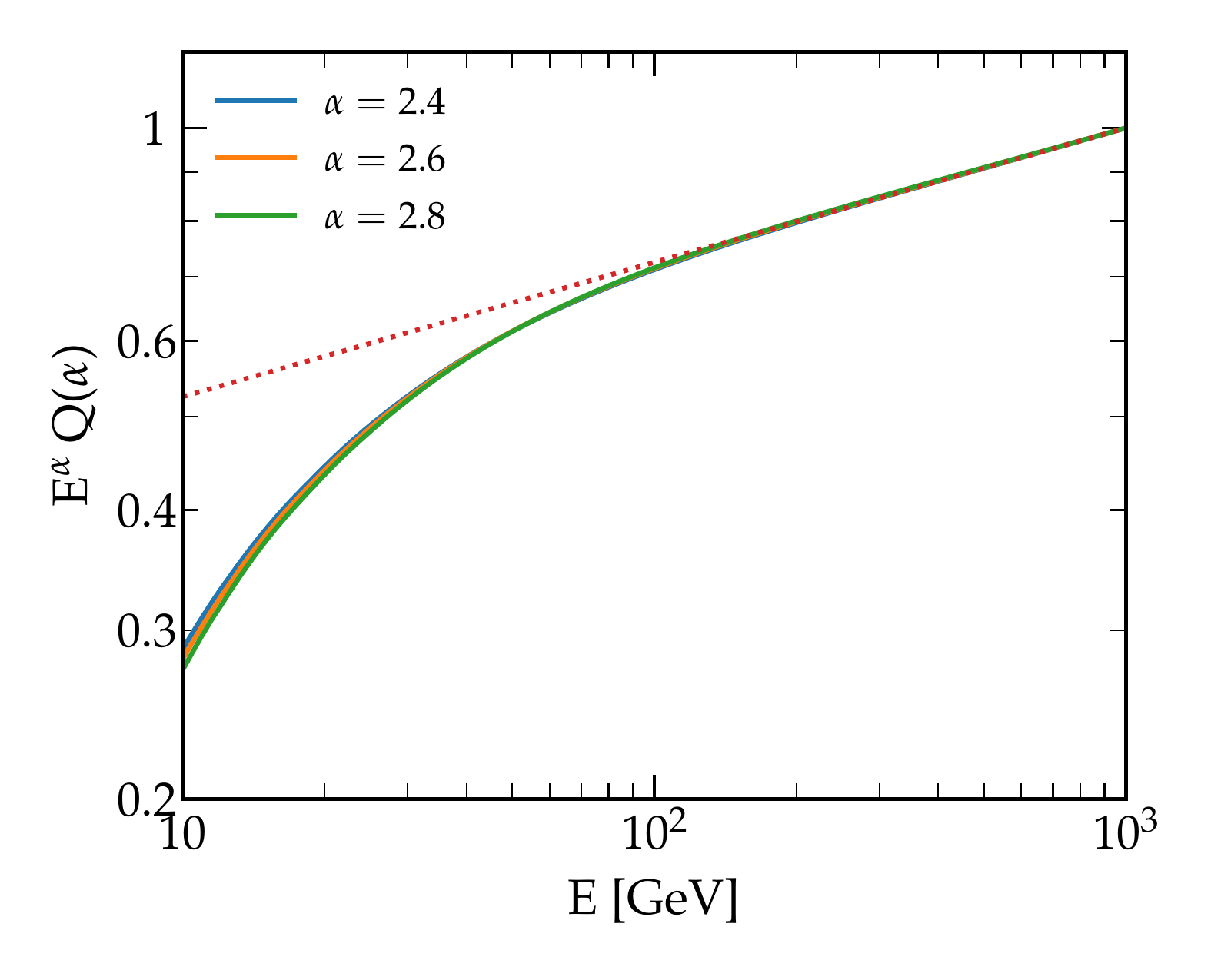}
\caption{Energy dependence of the antiproton source function \(Q(\alpha)\), multiplied by \(E^\alpha\) to emphasize deviations from the primary energy dependence, for different spectral indices \(\alpha\) of the primary proton parents. The curves represent calculations of \(Q\) for \(\alpha\) values of 2.4 (blue line), 2.6 (red line), and 2.8 (green line), demonstrating the weak sensitivity of \(Q\) to changes in \(\alpha\). Additionally, the dotted red line illustrates the behaviour of \(E^{0.15} \).}
\label{fig:apxsecs}
\end{figure}

One of the most appealing aspects of the NLBM has been the claim of predicting a roughly energy-independent antiproton/proton ratio, allegedly at odds with the predictions of the standard model of CR propagation. Here, we discuss the bases of this claim and the impact that cross sections of antiproton production have on it. 

If galactic sources inject protons with a power-law spectrum \(E^{-\alpha}\) and their residence time in the Galaxy depends on energy as \(\tau_{\rm esc} \propto E^{-\delta}\), then the observed proton spectrum should scale as \(E^{-(\alpha + \delta)}\). Meanwhile, secondary particles (e.g., boron nuclei or antiprotons), produced by collisions of CRs with the interstellar medium, should have a spectrum \(\propto E^{-(p + 2\delta)}\), if the production cross sections are independent of energy. Consequently, any secondary-to-primary ratio, such as boron over carbon (B/C) or antiproton over proton, should behave like \(E^{-\delta}\). 
While this trend is confirmed for B/C and similar ratios, for $\bar p/p$ below 300 GeV, this ratio is compatible with being constant~\cite{AMS:2016oqu,AMS:2021nhj}. 
Based on this line of thought, in the NLBM the escape time is assumed to be energy independent, \(\delta \sim 0\).
%an assumption that turns out to be central to the NLBM. 

A major development in this field has been the recent re-evaluation of the antiproton production cross sections (see, e.g.,~\cite{diMauro:2014zea,Korsmeier:2018gcy}), as based on better measurements and better modeling of the results of such measurements. These studies have revealed that the production cross section of antiprotons increases slightly with energy. 

As an illustrative example, Fig.~\ref{fig:apxsecs} shows the antiproton emissivity obtained by assuming a single power law for the parent (primary) proton distribution with slope $\alpha$, multiplied by $E^\alpha$. For an energy-independent cross section, this quantity would be flat. The figure demonstrates that, nearly independently of the value of $\alpha$, the spectrum of secondary antiprotons increases as $E^{\beta_\sigma}$, with \(\beta_\sigma \sim 0.15\) describing the energy dependence of the $\bar p$ production cross section. 

Furthermore, since antiprotons carry roughly 5-10\% of the energy of the parent protons, the observed spectrum effectively samples the proton spectrum at ten times higher energies, where it is harder than at low energies due to the spectral break, see sec.~\ref{sec:p+He}. Taking these two effects into account, the equilibrium antiproton spectrum can be written as
\begin{equation}
N_{\bar{p}} \;\propto\; \frac{E^{-\alpha_{\rm p}^\prime} \, E^{\beta_\sigma}}{E^{\delta}}~,
\end{equation}
where \( \alpha_{\rm p}^\prime \) is the parent proton spectrum above the break.

This implies that the ratio of antiprotons to protons at energies below the break ($\sim 300-500\,$GeV) is
\begin{equation}
\frac{N_{\bar{p}}}{N_p} 
\;\propto\; \frac{E^{-\alpha_{\rm p}^\prime} \, E^{\beta_\sigma} \, E^{-\delta}}{E^{-\alpha_{\rm p}}}
\; \propto \; E^{\Delta \alpha} \, E^{\beta_\sigma} \, E^{-\delta} \propto E^{0.35 - \delta},
%\;\;\longrightarrow\;\; {\color{red}0 = 0.2 + 0.14 - \delta}.
\end{equation}
where \( \Delta \alpha \sim 0.2 \) is the difference between the proton slope above and below the break (see section~\ref{sec:p+He}).
 
Hence, once the latest antiproton production cross sections are taken into account, the requirement that the antiproton-to-proton ratio remains nearly constant forces the the galactic residence time to decrease with energy with a slope \(\delta \approx 0.35\), consistent with measurements of the B/C ratio. This outcome aligns well with the extensive literature that demonstrates that the halo model grammage naturally reproduces the normalization and the shape of the antiproton flux~\cite{Boudaud:2021oro,Lv:2023gdt,Silver:2024ero,DelaTorreLuque:2024ozf}.

We conclude that the observed approximate energy independence of the antiproton-to-proton ratio is incompatible with an energy-independent Galactic grammage, i.e., \(\delta \sim 0\), namely it is, after all, incompatible with one of the most critical assumptions of the NLBM. 

\section{Conclusions}
\label{sec:conclude}

The undisputed success of the standard model of Galactic CR transport, developed over a few decades since the early 1960's and continuously fed by new measurements of the spectra of primary and secondary nuclei, as well as antimatter, has recently been questioned, mainly because of the discovery of the anomalous flux of CR positrons. While it is understood that there are sources of primary positrons in the Galaxy, most notably pulsar wind nebulae~\cite{Hooper:2008kg,Fermi-LAT:2009ppq,Delahaye:2010ji,Blasi:2010de,DiMauro:2014iia,Evoli:2020szd}, that may account for the anomaly, it has been proposed that positrons might be of pure secondary nature~\cite{Blum:2013zsa,Lipari:2016vqk,Lipari:2021edz} if the model of CR transport in the Galaxy is substantially modified. The NLBM provided the framework for such modifications~\cite{Cowsik:1973yd,Cowsik:1975tj} and seemed to receive additional support by the measurement of the flat energy dependence of the ratios $\bar p/p$ and $e^+/p$~\cite{Cowsik:2016wso,Cowsik:2022eik}. In particular, the latter lent foundation to the idea that, contrary to the common wisdom in which CRs are diffusively confined in an energy dependent manner in an extended magnetized halo, the galactic confinement region should be small and CRs should escape in an energy independent way. This setup is required in order to make energy losses of positrons negligible up to $\sim 1$ TeV. 

In addition, the recent discovery of regions of suppressed diffusivity around selected PWNe, the so called TeV halos \cite{HAWC:2017kbo,LHAASO:2021crt}, seemed to be lending support to the idea of cocoons near the sources, where in principle CRs could accumulate additional grammage. 

In the present work, we critically discuss all  ingredients and assumptions of the NLBM and test them against the wealth of data that are now available on the fluxes of primary and secondary nuclei, unstable isotopes and antimatter. 

In terms of primary nuclei, we point out that in the context of the NLBM it is impossible to attribute the observed spectral hardening measured at a rigidity of few hundred GV to transport processes, as it is done in the standard model. This is particularly disturbing since this hardening seems to occur at the same energy where secondary/primary ratios show a feature, that in the NLBM is to be attributed to the transition from grammage accumulated in the cocoons to Galactic grammage. This coincidence of numerical values, in the NLBM is assumed to be fortuitous and due to unknown phenomena taking place inside the sources of CRs. In addition, since in the NLBM, Galactic transport is energy independent, it follows that the observed  spectra of primary nuclei is bound to be the same as the source spectrum. Hence the source spectra must be very steep (unlike the ones expected based on DSA) and broken. Moreover, since leptons suffer little losses at $E\lesssim 1$ TeV, this must be true for primary electrons as well, which implies that the electron source spectrum must be $\propto E^{-3.4}$, again very steep compared with the standard predictions of DSA.

We have shown that the assumption of the NLBM of an energy-independent escape time from the Galaxy is in contradiction with both the spectral shape and the normalization of the positron and antiproton spectra.
In particular, the energy-independent Galactic grammage needed to explain the B/C ratio fails to reproduce the normalization of the antimatter fluxes. Hence, unlike the halo model, the NLBM is incapable of providing a self-consistent picture of CR transport.
The finding is further exacerbated looking at the slope of these spectra.

We demonstrated that the energy independence of the measured $\bar p/p$ spectrum, often listed as a strong prediction of the NLBM, is, in fact, incompatible with the NLBM once the most recent parameterizations of the cross sections are adopted and the observed spectral break in H and He is included in the model. The combination of these two effects lead us to conclude that the confinement time in the Galaxy needs to decrease with energy, roughly as $E^{-0.35}$, at odds with the NLBM and somehow perfectly in line with the standard model of CR transport. 

Finally, we show that the preliminary data on the $^{10}$Be/$^9$Be ratio, traditionally considered as one of the best indicators of the CR confinement time in the Galaxy, cannot be explained by the NLBM once the spectra of primary and secondary stable nuclei are fixed. In particular, preliminary measurements by AMS-02~\cite{Dimiccoli:2023dbx} require that the NLBM adopts a much longer confinement time in the Galaxy than allowed by other observables. 

We conclude that the version of the NLBM that is typically used in the literature is actually incompatible with data, and at best requires substantial revisions. Interestingly however, such modifications tend to make the NLBM closer to the standard model of CR transport. Despite this negative conclusion, we want to stress that with the level of accuracy that current measurements reached, it is clear that subdominant effects, due to grammage accumulated by CRs inside or around their sources, at some point must appear in the data and would be of great importance to clarify the global picture of the origin of Galactic CRs~\cite{DAngelo:2015cfw,Evoli:2019wwu,Recchia:2021vfw,Jacobs:2021qvh}.   

\acknowledgments

This work has been partially funded by the European Union – Next Generation EU, through PRIN-MUR 2022TJW4EJ and by the European Union – NextGenerationEU under the MUR National Innovation Ecosystem grant ECS00000041 – VITALITY/ASTRA – CUP D13C21000430001. CE and PB are also supported by the research project TAsP (Theoretical Astroparticle Physics) funded by the Istituto Nazionale di Fisica Nucleare (INFN).
The work of BS was supported by NSF through grants NSF-2009326, and NSF-2010240.
\bibliographystyle{apsrev4-2}
%\bibliography{2025-nlbm}

%\newpage

%\appendix

%\input{apthreshold}

\end{document}